\documentclass{cernatsnote}
\usepackage[colorinlistoftodos]{todonotes}
\usepackage{placeins}
\usepackage{siunitx}
\usepackage{enumerate}
\usepackage{lscape}
\usepackage{rotating}
\usepackage{float}
\usepackage{caption}
\usepackage{url}
\usepackage[switch]{lineno}
\pdfoutput=1

\documentlabel{CERN-SPSC-2022-009}

\title{Study of alternative locations for the SPS Beam Dump Facility}
\author{
	Oliver~Aberle, Claudia~Ahdida, Pablo~Arrutia, Kincso~Balazs, Johannes~Bernhard, Markus~Brugger, Marco~Calviani, Yann~Dutheil, Rui~Franqueira~Ximenes, Matthew~Fraser, Frederic~Galleazzi, Simone~Gilardoni, Jean-Louis~Grenard, Tina~Griesemer, Richard~Jacobsson, Verena~Kain, Damien~Lafarge, Simon~Marsh, Jose~Maria~Martin~Ruiz, Ramiro~Francisco~Mena~Andrade, Yvon~Muttoni, Angel~Navascues~Cornago, Pierre~Ninin, John~Osborne,  Rebecca~Ramjiawan, Pablo~Santos~Diaz, Francisco~Sanchez~Galan, Heinz~Vincke, Pavol~Vojtyla \; \\	
    
	\center CERN, CH-1211 Geneva, Switzerland
}
\date{1 March 2022}
\begin{document}
\maketitle

\begin{abstract}

As part of the main focus of the BDF Working Group in 2021, this document reports on the study of alternative locations and possible optimisation that may accompany the reuse of existing facilities with the aim of significantly reducing the costs of the facility. Building on the BDF/SHiP Comprehensive Design Study (CDS), the assessment rests on the generic requirements and constraints that allow preserving the physics reach of the facility by making use of the $4\times 10^{19}$ protons per year at 400\,GeV that are currently not exploited at the SPS and for which no existing facility is compatible. The options considered involve the underground areas TCC4, TNC, and ECN3. Recent improvements of the BDF design at the current location (referred to as `TT90-TCC9-ECN4') are also mentioned together with ideas for yet further improvements. The assessments of the alternative locations compiled the large amount of information that is already available together with a set of conceptual studies that were performed during 2021.

The document concludes with a qualitative comparison of the options, summarising the associated benefits and challenges of each option, such that a recommendation can be made about which location is to be pursued. The most critical location-specific studies required to specify the implementation and cost for each option are identified so that the detailed investigation of the retained option can be completed before the end of 2022.
\end{abstract}
\\ \\ \\ 

\begingroup
\color{black}
\tableofcontents
\endgroup

\pagebreak
\section{Context and objective}

An Expression of Interest~\cite{Bonivento:1606085} for a new experimental facility to search for hidden sector particles was submitted to the SPSC in October 2013. One of the underpinning goals of the initiative was to identify the physics potential of the $4\times 10^{19}$ protons per year at 400 GeV that SPS delivered to CNGS up to 2012 but that are still unexploited and for which no present facility at the SPS is compatible. A first location study was performed for the EoI. It considered TCC4, TNC, ECN3, and TT61. As these facilities were hosting projects recently approved at the time (AWAKE, HiRadMat, and NA62, respectively) and TT61 was disfavoured for environmental reasons, the studies continued with a focus on an entirely new construction~\cite{TaskForce2014}. However, the development of the concepts for the new facility followed a strategy that makes a large part of the studies generic to several locations around the SPS. 

Following the favourable SPSC recommendations on the EoI, a group of institutions from CERN Member and non-Member States and CERN formed the SHiP Collaboration and proceeded with the preparation of a Technical Proposal (TP)~\cite{Anelli:2007512,SHiP:2015gkj} on the detectors and the proposed SPS experimental facility~\cite{Arduini:2063023, Goddard:2063299, Calviani:2063300, Strabel:2063301, Osborne:2063302, Ahdida_2019, Bartosik:2650722}. The documents for the TP and an extensive report on the Physics Case~\cite{2016SHiPPhysicsCase} were submitted together to the SPSC in 2015. 

The review of the TP concluded with a recommendation to proceed with a three-year Comprehensive Design Study, under the auspices of the CERN Physics Beyond Colliders (PBC) initiative, with the goal of submitting a proposal for the Beam Dump Facility (BDF)~\cite{Ahdida:2703984} and the experiment (SHiP)~\cite{Ahdida:2654870,Ahdida:2704147} to the 2020 European Particle Physics Strategy Update.  More details on the physics sensitivity of the experiment can be found in Ref.~\cite{SHiP:2018xqw,Ahdida:2020new,SHiP:2020noy}.

The Deliberation Document of the 2020 Update of the European Strategy for Particle Physics~\cite{European:2720131} recognised the BDF/SHiP proposal as one of the front-runners among the new facilities investigated within the PBC studies. With regards to the cost of the CDS design of the facility, the project could not, as of 2020, be recommended for construction considering the overall recommendations of the Strategy.

In line with the ESPP recommendations and in view of the importance of the CERN injector complex as a provider of physics programmes complementary to CERN's primary large-scale research facility, and the strong motivation behind the BDF facility, a continued programme of R\&D was launched in 2020~\cite{MTP2020}. In the new mandate~\cite{BDFmandate}, the BDF Working Group has been tasked with reviewing the design of the beam delivery, target design, and suppression of beam induced background, as well as reviewing the layout and most suitable site at CERN, including reuse of existing facilities. 

The renewed effort has initially focused on a broad location study and re-optimisation of key components with the goal of reducing costs, while preserving the original physics scope and reach of the facility. This also means preserving the main goal of constructing a facility that is capable of coping with the $4\times10^{19}$ protons per year at 400\,GeV, while still maintaining the capabilities of the other existing experimental facilities at the SPS. This effort is accompanied by a re-optimisation by the SHiP collaboration of the layout and of the key components of the experiment with the goal of reducing in size the overall volume required for the experiment, allowing integration into the alternative areas after limited modifications. A formal collaboration agreement with the SHiP institutes has been established through a Memorandum of Understanding~\cite{EDMS_MoU} that ensures a coherent optimisation effort between the facility and the experiment.

\begin{figure}[h]
    \centering
    \includegraphics[width=0.80\textwidth]{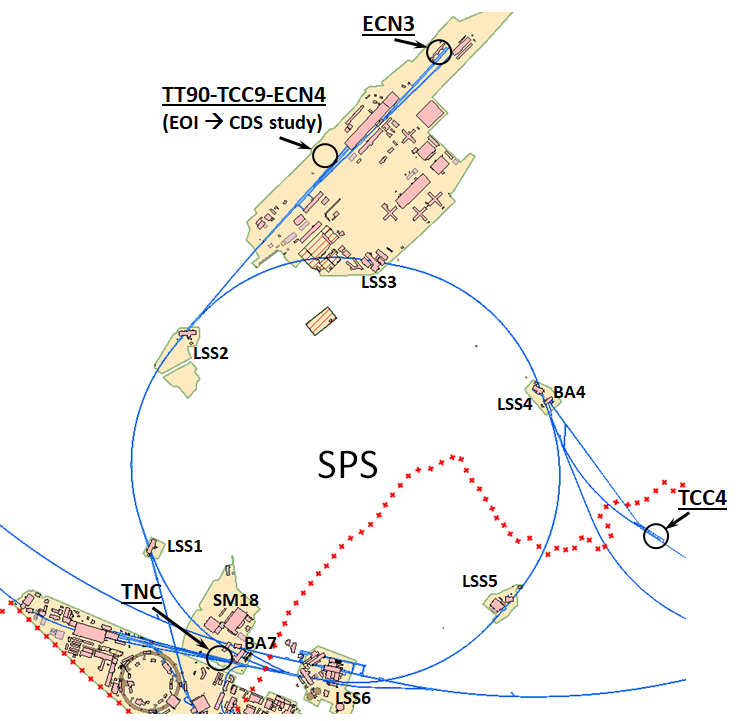}
    \caption{Overview of the locations considered for the implementation of the BDF.}
    \label{fig:LLO_overview}
\end{figure}

This note reports on the assessment of three alternative locations for BDF (see Fig.~\ref{fig:LLO_overview}). The study is largely based on available information together with a few new key conceptual studies performed to identify the benefits and the challenges associated with each location. The note concludes with a comparative summary of the main aspects.
The report also suggests the principal studies necessary to further develop the most promising option on the time-scale of one year.

The document aims to be an aid in the decision making concerning the option that should be retained for the preparation of the final design of BDF, and concerning the strategy and priorities for the CERN experimental programme.

Appendix II has a glossary for the acronyms used in this document.

\section{BDF generic requirements\nopagebreak}
\label{sec:requirements}

The main purpose of the BDF is the search for feebly interacting particles, including hidden sectors and light dark matter. These are predominantly accessible through the decays of heavy hadrons and radiative processes. The production mechanisms and the currently unexplored ranges in mass and coupling make the SPS 400\,GeV high-intensity beam unique. The wide range of physics modes also largely shares phenomenological characteristics, making it possible to design a general-purpose hidden sector facility based on a common optimisation. As a by-product, the setup also becomes suitable for studying interactions of neutrinos, most interestingly of tau neutrinos.

The design of the facility at the SPS is based on returning to the full exploitation of the CERN accelerator complex with the SPS at its present performance. The SPS serves the LHC (HL-LHC) and a set of fixed-target facilities for physics and for R\&D, together using up to $1.5\times10^{19}$ protons per year. Since the decommissioning of CNGS, up to $4\times10^{19}$ protons per year at 400\,GeV have been left available but unexploited~\cite{Bartosik:2650722}. At a nominal spill intensity of $4\times 10^{13}$ protons, up to $2\times 10^{20}$ protons on target could be delivered to the new facility in about five years of operation, while respecting the requirements of the HL-LHC operation as well as the existing facilities. The recent upgrades of the SPS may provide it with the capability of delivering more than $4\times10^{13}$ protons per spill and have the potential of providing the BDF with an additional $2\times10^{19}$ protons per year. This could also give more flexibility for programming the CERN supercycle and duty cycle of other facilities.

Furthermore, the experiment relies on the unique feature of slow beam extraction over a timescale of around a second to control the background from random combinations of muons producing fake decay vertices in the decay volume. In the CDS design with a solid production target, the slow extraction is also required to dilute the large beam power deposited on the proton dump both spatially and temporally.

The required intensity and one second spills necessitate dedicated SPS cycles for BDF. Where North Area locations are considered, a dedicated optics in the relevant transfer lines will be required because the primary proton beam should not undergo splitting before delivery to the BDF target complex.

In order to maximise the production of heavy flavoured hadrons, and photons, and at the same time provide the cleanest possible background by suppressing decays of pions and kaons, the target should be a $>10\,\lambda$ thick combination of materials with the highest possible atomic mass and number. The beam power associated with ${\cal{O}}(10^{19})$ protons at 400\,GeV per year makes it invariably necessary to envisage internal cooling of the target and the downstream absorber, and a proximity shielding embedded in a cooled inert gas to reduce activation and radiation-induced corrosion. The induced activation requires a target complex compatible with remote handling and adequate shielding to avoid activation, or further activation, of surrounding soil, beam-line components upstream, and the downstream experimental area. The constraints on the target and the target complex are in these aspects independent of the purpose of the beam.

The beam induced background flux is manageable with a 5\,m hadron absorber and a muon shield system of about 25\,m in length. A 50\,m long decay volume under vacuum eliminates background events from neutrino interactions in the fiducial volume. In this configuration the re-optimised version of the SHiP detector preserves the large acceptance of the CDS design, occupying in total a space of about 80\,m in length, and a width and height of about $6\times 11$\,m$^2$. The latter requirement comes from the size of the spectrometer section and the particle identification detectors that extend over the last 18\,m. New preliminary studies of the background from muons interacting with the cavern walls indicate that the distance to the cavern walls in the current design may also be relaxed at the cost of a moderate but manageable increase of background rates~\cite{Ferro-Luzzi:2799389}.

The entire facility is required to be situated below ground to contain the radiation around the target complex, and to range out the muon flux escaping the experimental area. In order to consider the facility as fully optimised, the dose to members of the public must remain below 10\,$\mu$Sv/year~\cite{SafetyCodeF}. Since the muon deﬂection is done in the horizontal plane and the vast majority of high-energy muons are well contained underground, the required attenuation is achieved with about 650\,m of ground downstream of the facility.

A dedicated surface building is required to house the target systems ancillaries, such as the cooling and ventilation systems as well as control racks. Dedicated space and handling equipment are also required to store activated components that may come from the operation of the facility. Depending on the location, additional space in surface buildings may also be needed to house power converters for the beam line and the experiment. Synergies with existing or new CERN central facilities are to be envisaged and pursued.

Operation of the experiment requires an adjacent surface building for services, electronics racks, and control room. The experiment installation phase needs access to an assembly hall with height and width similar to the surface hall in the CDS design of the TT90-TCC9-ECN4 option over a period of 2\,-\,3 years, not necessarily adjacent to the facility but with an adequate road connection for transports.


\section{CDS design TT90-TCC9-ECN4}

\subsection{Brief description}

CERN's North Area has a large area on the Jura side of TDC2-TCC2 (Fig.~\ref{fig:LLO_overview}) which is for the most part free of structures and underground galleries, and which could accommodate the BDF. In this configuration, the BDF relies on the slow extraction of 400\,GeV protons from Long Straight Section 2 (LSS2). The technology used in LSS2, including the five high-field ($\sim$ 10 MV/m) electrostatic septa, has remained largely unchanged since the conception of the SPS in the 1970s. In recent years, operational problems linked to the activation of LSS2 from the increasing proton flux requested by the North Area, along with the prospect of the BDF demanding an even higher flux in the future, have motivated a significant R\&D effort to improve the slow extraction efficiency~\cite{Balhan:2668989}. A range of different state-of-the-art techniques have been prototyped, validated with beam and implemented operationally in the last five years. The different techniques promise up to a factor of ten reduction in the extraction inefficiency. In addition, improvements in the choice of material, and the application of local shielding, would further improve the radiological situation and permit the extraction of higher proton fluxes to the North Area. Although the R\&D is still in active development, the factor of 4-5 reduction in beam loss per extracted proton needed to keep today's level of radioactivity in LSS2 during BDF operation looks to be within reach. The bent silicon crystal technology used to shadow the blade of the electrostatic septa in LSS2 has already demonstrated a factor of two loss reduction and was deployed on the operational beam to the North Area throughout the 2021 operational year. An optimised crystal shadowing system has been installed in LSS4 for tests with beam in 2022. It is predicted to reduce the beam loss in the LSS2 by a factor of 4 - 10, depending on the particular choice of crystal technology employed.

After about 600\,m of the present TT20 transfer line, a new splitter/switch magnet system, which will maintain compatibility with the existing North Area operation, deflects the beam into a dedicated new transfer line (TT90) for the BDF. The new transfer line is connected to the existing tunnel by a modified junction cavern. The line consists mainly of dipole magnets necessary to bend the beam away from the TDC2-TCC2. 

The beam is dumped on a 12$\lambda$ target/dump composed of disks of molybdenum alloy and tungsten, cladded with a diffusion bonded tantalum alloy~\cite{PhysRevAccelBeams.22.113001,PhysRevAccelBeams.22.123001}. The disks are separated by thin gaps for active water cooling within a closed pressurised vessel. In order to maintain the beam-induced thermo-mechanical stresses below acceptable limits, the beam is swept over the target in a circular fashion with the help of orthogonal kicker magnets together with $\sim 100$\,m drift space on the beam line in order to dilute the energy deposited. 

The target system is housed in a purpose-built target complex~\cite{Kershaw_2018}, with the target located within a bunker at around 15\,m below the ground level. To minimise the risk of ground activation and groundwater contamination at this new site, it has been found that the specific activation of dissoluble radionuclides in the ground may be kept under control if the prompt dose rates are kept below 1\,mSv/h. As a result, the target system is surrounded by approximately 3700 tonnes of cast iron and steel shielding with outer dimensions of around $6.8 \times 7.9 \times 11.2$\,m$^3$, including the downstream hadron stopper. The first layer of shielding around the target is water cooled to deal with the heat generated by the particle showers. The target and the complete shielding is enclosed in a vessel filled with helium as the choice of inert gas to reduce air activation and reduce radiation-induced corrosion.

The target complex has been designed to house the target bunker, along with the cooling, ventilation and helium purification services below ground level~\cite{Kershaw_2018,Avigni_2019}. Remote handling for manipulation of the target and surrounding shielding is mandatory due to the high residual dose rates expected after operation.  The design allows for removal and temporary storage of the target and shielding blocks in the cool-down area below ground level and includes dedicated shielded pits for storage of the equipment with the highest dose rates.

In order to host the muon shield and the SHiP detectors, the underground hall is 20\,m wide and 120\,m in length. A 100\,m~$\times$~26\,m surface hall is located on top of the underground hall. The original installation plan foresees parallel pre-assembly in the surface hall in three principal work zones with the help of a 40-tonne and a 10-tonne crane, in parallel to final assembly in the underground hall using a dual 40-tonne crane. Three 14.5\,m~$\times$~18\,m access openings between the surface hall and the underground hall provide direct access to the principal detector installation areas. For shielding purposes, each opening is covered by concrete beams during operation. A large service building, adjacent to the surface hall, houses all services related to the infrastructure and the detector, control room, etc.

For the radiation protection studies, the soil surrounding the facility was modelled with a conservative density and ground profile. In this way, location-dependent density differences, and lower densities as a result of civil engineering works, are conservatively accounted for. The area downstream of the ECN4 is furthermore planned to be covered with 6\,m of the remaining soil from the excavations, bringing the dose rate down below the limit of 0.5~$\mu$Sv/h for a Non-Designated Area~\cite{RP_zoning}.

Table~\ref{tab:cost-summary-material-overview} reports the high-level summary of estimated material costs for the CDS design of the BDF constructed entirely with no reuse of existing infrastructure, reproduced here for reference.

\begin{table}[htb]
\begin{center}
\caption{Overview summary of the material cost estimates for the CDS design of the BDF in the TT90-TCC9-ECN4 option. Reproduced from the 2019 BDF CDS report~\cite{Ahdida:2703984}\\}
\label{tab:cost-summary-material-overview}
\begin{tabular}{lrr}
\hline
\textbf{Work Package/System} & \multicolumn{2}{l}{ \textbf{Estimate [MCHF]}}      \\
\hline
Civil Engineering                               & & 68.2 \\
\,\,\,\,\,\,\,\,WP1 - Experimental area        &  31.7 & \\
\,\,\,\,\,\,\,\,WP2 - Target complex       &  16.5 & \\
\,\,\,\,\,\,\,\,WP3 - 85\,m extraction tunnel + associated buildings       &  10.9 &  \\
\,\,\,\,\,\,\,\,WP4 - Junction cavern + 79\,m extraction tunnel        &  7.3 & \\
\,\,\,\,\,\,\,\,WP5 - Site investigations + roads and site preparation       &  1.9 & \\
Beam-line and junction cavern hardware     &  & 9.6  \\
Target and Target Complex infrastructure  & &  45.5  \\
Cooling and ventilation  & &  13.7  \\  
Electrical distribution  & &   5.6 \\
Survey and alignment  & & 1.1   \\
Access, safety, RP, controls & &   6.0 \\
Transport (inc. cranes and lifts)  & &  5.1 \\
(De)Installation & & 3.6 \\
\hline
Total & &  158.4 \\
\hline 
\end{tabular}
\end{center}
\end{table}

The civil engineering costs represent a significant proportion of the overall cost in the CDS design. In addition, it is considered to have an inaccuracy -20\% to +40\% until the project is further developed. The proximity of existing infrastructures such as the TCC2 cavern and the transfer tunnels TT81, 82 and 83 requires continuous monitoring of the construction-induced vibration and careful planning of the works due to the activation of the tunnels and the surrounding earth. For the construction of the junction cavern, a relevant part of the existing TDC2 must be demolished and the existing machine and services need to be removed prior to the works. Handling the demolished concrete and the activated earth will require special procedures even if it will be reused as back fill with the most activated soil placed closest to the new junction cavern and extraction tunnel, to avoid disposal off-site and additional soil activation in the future around the BDF. Diaphragm walls have been foreseen to ease the construction close to the existing infrastructures, to reduce the volume of the excavated earth, and to prevent potential groundwater contamination in the areas with high groundwater level and high permeability of the earth layers.

\subsection{Possible optimisations of the TT90-TCC9-ECN4 option}
\label{sec:TT90optim}

Several possibilities to optimise and reduce the cost and risks of the TT90-TCC9-ECN4 option have been identified.

In the CDS design, the slow extracted beam from the SPS is switched towards the new TT90 beam line using an upgraded laminated splitter magnet capable of cycle-to-cycle field reversal. However, faced with the complexity and estimated cost of this solution, several different arrangements of the splitter region have been investigated. In particular, the study revealed the possibility of switching the beam to the BDF above the existing splitters, with minimal changes to the area and no impact on the nominal split beam sent to the North Area. This change not only reduces the overall cost of switching the beam towards the new TT90 beam line, but also reduces the risk involved in the R\&D still needed to realise the laminated splitter and minimises perturbation to nominal North Area operation.

\begin{figure}[!htpb]
\centering
\includegraphics[width=0.9\linewidth]{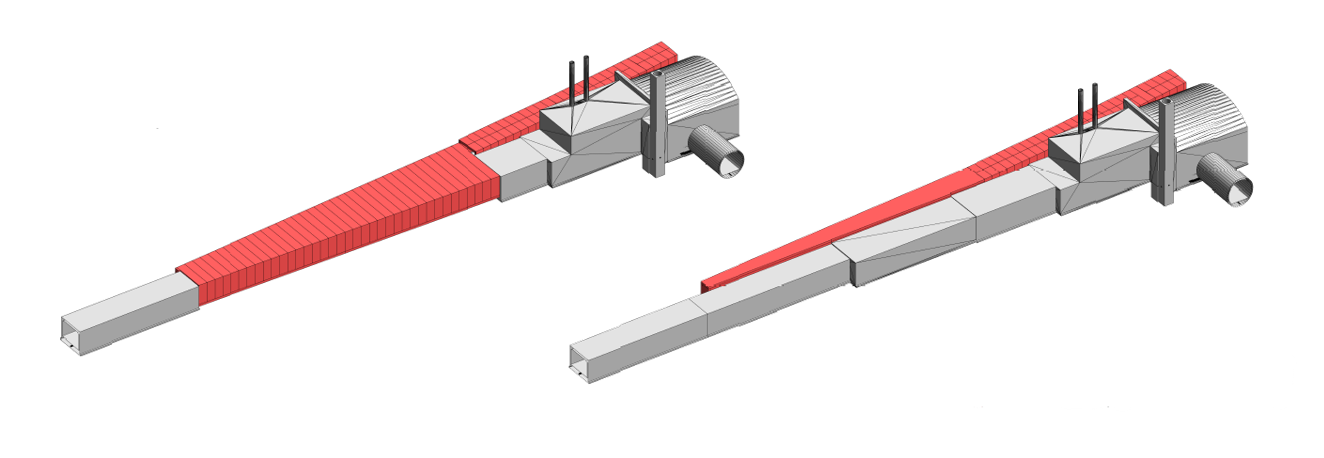}
\caption{3D rendering of the 2018 CDS proposal on the left and the optimised solution on the right with existing structures in grey and new ones in red in the TDC2 area.}
\label{fig:TT90:wall_penetration}
\end{figure}

In the CDS design, the existing TDC2 cavern and the new extraction tunnel are connected through a new 75\,m long junction cavern. The construction involves important civil engineering works such as the demolition of the existing tunnel along the entire length of the new cavern. As a result, complete removal and re-installation of all systems and services in the affected area are necessary. To reduce the impact on the existing infrastructures and the related risks, a new solution has been proposed to create only a small penetration through the wall of the TDC2 cavern to allow the beam line to pass through into the newly constructed connection tunnel with TT90 (Fig.~\ref{fig:TT90:wall_penetration},\ref{fig:TT90_option_Details},\ref{fig:TT90_Wall_penetration_option_Plan_view} in Appendix I). The works may be carried out with most of the existing TDC2 equipment and services remaining in-situ during the construction phase. The external and internal assessment of this new option demonstrated its feasibility and important reduction both in cost and risk~\cite{balazs:wall_penetration}.

The earth retaining systems required throughout the facility during the civil works have also been re-assessed and optimised to further reduce the cost. Diaphragm walls were initially proposed along the whole length of the new extraction tunnel mainly to reduce the impact on the adjacent existing infrastructures, to significantly reduce the extent of the excavations required, and to act as a permanent barrier for the groundwater reaching the tunnel. As they incur significant costs, alternative solutions have been investigated as well. An open-cut excavation could be adopted instead of using diaphragm walls along the first 100\,m of the new TT90 tunnel and in the area of the junction with TDC2. The open cut option would require additional measures to protect the existing tunnel and surface assets due to unloading and temporary asymmetric soil loading but would still significantly lower the overall civil engineering cost. 

The target complex as an entirely new construction was designed with the constraint that the civil engineering infrastructure of the BDF must not be activated, hence simplifying the re-purposing of the installation in the future and simplifying the decommissioning in the long-term. Such a very stringent requirement is not imposed for any other similar infrastructures at CERN. Relaxing this requirement  would lead to a sizable reduction of the iron shielding and a simplification of the handling system, resulting in a significant cost reduction. These optimisations will further result in a reduced footprint, hence indirectly reducing the civil engineering costs. Cost reduction could also be derived from employing nitrogen as inert gas instead of helium. This will ease the vessel tightness requirement and largely simplify the purification system, as well as reducing operational costs. A dedicated assessment will be required.

The investigations of the alternative fully underground options have made it necessary to reconsider the detector assembly and installation procedures, and the detector access systems, also opening possibilities to review the design strategy for the TT90-TCC9-ECN4 option. A significant cost in the CDS design is associated with the civil engineering of the experimental area. The depth of the underground hall requires self-supported diaphragm walls on the perimeter to support high lateral loads. The re-optimised experiment layout makes it possible to reduce the size of the hall, resulting in shorter diaphragm walls and less reinforced concrete structures, providing a substantial cost reduction. Furthermore, the initial plan for the construction and installation of the experiment relies entirely on a dedicated surface assembly hall located on top of the underground hall. The assumed parallel assembly activities require a hall covering the full length and width of the underground experimental hall and using three independent assembly areas and flexible access openings between the surface and the final detector location. Hence, the surface hall is currently designed as a conventional clad steel frame structure, the bottom slab having three openings to allow the vertical access to the underground hall filled with removable, precast, pre-stressed and pre-tensioned RC beams. Alternative existing assembly halls with sufficient space are available on the North Area site. Re-optimisation of TT90-TCC9-ECN4 includes reducing the size of the surface hall to only support intermediate storage and assembly, and a single access opening to the underground area. This lowers the volume of the steel works required and the number of the removable RC beams, further reducing the overall civil engineering cost.

\section{Alternative locations}

\subsection{Overview of options}

The study of alternative locations for the BDF has focused on three existing underground facilities, indicated in Fig.~\ref{fig:LLO_overview}. 

The {\bf ECN3} experimental area was constructed at the end of the 1970s as part of the SPS North Area High-Intensity Facility for fixed-target experiments~\cite{Brianti:643069}. It was commissioned in 1980 and hosted a long series of experiments starting with NA10. The ECN3 experimental hall together with the TCC8 target hall are located at the end of the 750\,m long TT83-TT85-TDC85 transfer tunnel from TCC2. ECN3 is about 100\,m in length, 16\,m wide, and has a free height below the 45-tonne crane of 8.9\,m along its entire length. TCC8 and ECN3 have no slope. Both are located entirely below the natural ground level with a 8\,m thick additional layer of earth added on top. The 170\,m long and 10\,m wide TCC8 target hall was built for free-standing target systems in direct connection with ECN3, and is equipped with a 30-tonne crane with 4.5\,m free space below the hook. ECN3 currently hosts the NA62 experiment~\cite{Ceccucci:832885, Collaboration:2691873}.

{\bf TNC} was the former location of the T9 target, horn, reflectors, and helium vessels of the West Area Neutrino Facility that was constructed in the early 1970s and operated between 1976 and 1998. The hall is about 160\,m in length and is located on the TT60-TT66 transfer line from the SPS LSS6 at a depth of about 30\,m just after TCC6 and the SPS BA7 access point (TJ7). It has an upward slope of 4.6\%, and a full width and height of 6\,m\,$\times$\,3.6\,m. In addition to the BA7 access, the TNC hall has a personnel escape shaft at the rear end. A large scale decommissioning was performed in 2010 - 2012 to recast the area for the high-energy, high-intensity general-purpose irradiation facility HiRadMat that uses the first 60\,m of TNC. HiRadMat has been in operation since 2012~\cite{Harden_2019}.

{\bf TCC4} is the target hall of the former CERN to Gran Sasso (CNGS) facility. The hall is about 120\,m in length and is located at a depth of about 60\,m at the end of the TT41 transfer line branching off from the SPS in LSS4. The existing hall has a downward slope of 5.7\%, and a usable width of 5.6\,m and a height of 3.6\,m below the existing 7.5-tonne crane along the entire length. The hall is only accessible via the 800\,m long TAG41 from the SPS BA4. The CNGS facility was constructed in the years 2000 - 2005 and was in operation until 2012. While the downstream 100\,m of the hall with the CNGS target and the magnetic horns have been sealed off for cool-down, the upstream section currently hosts the AWAKE plasma acceleration R\&D project~\cite{awake}.

\subsection{Considerations}

The assessments of the different options considered the following aspects:

\begin{itemize}
\setlength\itemsep{-0.2em}
    \item Status of current projects hosted in the facilities
    \item Decommissioning (prerequisite): interference with existing installations
    \item Decommissioning (prerequisite): removal of existing infrastructure
    \item Impact on operation of adjacent facilities
    
    \item Beam delivery: energy/intensity
    \item Beam delivery: slow extraction
    \item Beam delivery: branch-off
    \item Beam delivery: transfer line routing
    \item Beam delivery: beam dilution
    
    \item Target system: shielding
    \item Target system: handling
    \item Target system: services and ancillaries
    
    \item Radiation protection: beam transfer losses
    \item Radiation protection: proximity to the surface, neighbouring facilities and the fenced and non-fenced CERN domain
    \item Radiation protection: residual dose rates
    \item Radiation protection: air and He/N$_2$ activation
    \item Radiation protection: soil activation
    \item Radiation protection: environmental impact
    
    \item Integration: availability of services (target, experiment)
    \item Integration: assembly halls
    \item Integration: installation
    \item Integration/civil engineering: geometry
    \item Integration/civil engineering: personnel/material shafts
    
    \item Civil engineering: interference existing neighbouring facilities
    \item Civil engineering: handling of activated concrete
    \item Civil engineering: construction techniques
    \item Civil engineering: access roads
    
    \item Safety: Personnel access beam-line
    \item Safety: Personnel access target complex
\end{itemize}

\section{ECN3 option}
\label{ECN3}

ECN3 is currently hosting the kaon experiment NA62~\cite{Ceccucci:832885, Collaboration:2691873}. The experiment was approved in 2008 and started physics data taking in 2016. The Research Board approved in June 2021 a last extension of the NA62 run until Long Shutdown 3~\cite{CERN-DG-RB-2021-505} in order to complete the physics programme.

\subsection{ECN3 - Beam delivery}

The ECN3 location in the North Area will profit from the existing slow extraction system and the R\&D effort already invested to improve it for the TT90-TCC9-ECN4 option.  

No additional beam line is needed for the BDF in ECN3. The beam will be transported from the SPS in TT20 and through both splitter magnets without splitting to maintain a high transmission efficiency and reduce losses to a minimum. The optics in TT20 will need re-matching for this purpose as was already designed and tested for the TT90-TCC9-ECN4 option. In the case of ECN3, the beam must be bumped up into the field-free hole in the yoke of the first splitter magnets before being bumped back down to pass though the dipole gap (high-field region) of the second splitter and deflected towards the T4 production target.

BDF in ECN3 will be served by dedicated SPS spills, as foreseen in the general specification of the BDF, without splitting in TDC2. A technical feasibility study will be needed to assess the possibility of cleanly transporting the beam through the North Area target system and to specify the cost and radiological impact of the required modifications to the T4 target area, including the T4 front-end and the TAX. After passing through the splitter magnets, the T4 target will be bypassed with the help of a vertical closed bump. The beam will then pass through the holes in the downstream TAX absorber and be transported to end of the P42 beam line. The function of the T4 production target for the H6 and H8 beam lines will be maintained on the SPS FT cycles. Pulse-to-Pulse Modulation operation of the system should be guaranteed so that the present operation of the North Area beam is left unperturbed. An alternative option would bypass the T6 target after deflection in the dipole gap of the first TT20 splitter in order to take the BDF spills via the old P6 channel into P42 with the help of a switching magnet in the M2 beam line.
  
Relatively minor modifications are expected in the rest of the P42 beam line to transport the BDF beam to the present location of the T10 target, which is $\sim$100\,m upstream of the foreseen location for the BDF target complex. Detailed modelling of accidental beam loss scenarios along the P42 bean line will need to be conducted to ensure that the SPS fast interlocking and the existing shielding are sufficient. The K12 beam line will be removed and replaced with the beam dilution system and a vacuum chamber extending to the BDF target complex.

The material access to the P42 beam line via TT84-EHN2, and the personnel access to TT85-TDC85-TCC8 via the existing shaft in building 912 are maintained.

\subsection{ECN3 - TCC8 Target complex}
\label{sec:ECN3_TC}

A first integration of the BDF experimental configuration suggests locating the centre of the target system approximately 30\,m upstream of the boundary between TCC8 and ECN3. 

The two existing target stations in TCC8 (T10 and T8) as well as the TAX and associated equipment should be dismantled in order to install the new beam line towards the new target area. T10 is currently in use for NA62, while T8 is not since several decades. Following similar past experience, the dismantling could be done in roughly 5-6 months following 6-8 months of cool-down, considering that the area activated by the two targets is limited to about 20\,m of TCC8. The dismantling can take place after the civil engineering required for the installation of the new facility, which is roughly 80\,m downstream of the current targets in TCC8.  

Following extensive analysis, comparison with similar installation worldwide, and taking into account the reuse of an already existing area, the shielding volume and configuration will be re-optimised profiting from the experience of the LIU SPS internal beam dump block project (TIDVG5), installed in the SPS during the Long Shutdown 2. TIDVG5 is designed to cope with roughly 300\,kW beam average power and a total of few 10$^{18}$~protons on target/y~\cite{PhysRevAccelBeams.24.043001,Pianese:IPAC2018-WEPMG004,TIDVG5_FUNCSPEC,TIDVG5_ENGSPEC}. In order to reduce air activation and radiation-induced corrosion, the target complex will be embedded in a nitrogen confinement. Due to the location in an underground cavern, the system could be simplified with respect to the baseline configuration employing helium as the T2K target system~\cite{ABE2011106}. Leak tightness is easier to achieve with nitrogen gas and results in a lower cost with respect to the helium configuration. A similar path has been taken by the LBNF project at Fermilab~\cite{osti_1366519}.

The development of the maintenance concept for the target station can directly profit from the extensive CDS studies carried out during 2017-2018 for the 'crane++' concept. In this concept, the handling is performed from the top with an overhead travelling crane but has the services standing outside of the shielding. The ECN3 option allows a simplification thanks to the installation in an existing larger underground cavern. As in all options, the maintenance scheme should be fully remote, potentially supported by telemanipulation and robotics tools.

The plan is to optimise the size of the shielding to fit the target station with its associated shielding to the current configuration of the TCC8 cavern. Minimal civil engineering works will be required at the proposed location of the BDF target complex to embed part of the shielding and services in the floor. The target complex will have to be embedded in a dedicated confined zone, separating it from the ECN3 hall and the rest of the TCC8. 

Fig.~\ref{fig:Target_complex_TCC8_integration} in Appendix I shows a preliminary implementation of the target complex as envisaged in the ECN3 option, taking into account the requirements from the target system, the water-cooled proximity shielding, as well as the magnetised hadron absorber. Similar designs are possible in the TNC and the TCC4 options. 

An auxiliary building will be required on surface to include all the dedicated services for the target complex.
Currently, there is no available space for this equipment in the close vicinity of TCC8/ECN3. A shielded area/building in order to separate, package and inspect radioactive components may be required on the surface.

The reuse of the current TCC8 overhead travelling crane for the target system is to be studied in detail. Depending on the complexity of the target complex it may be required to install on the same TCC8 crane rails a dedicated crane for the maintenance of the target complex as in the TT90-TCC9-ECN4 option.  

The current access shaft to ECN3 and TCC8 in building 911 may require enlargement. In this way it can be used to deliver all the services required for the target station. On the other hand, with a single shaft serving TCC8 and ECN3, the shared use for decommissioning of the current setup in TCC8 and in ECN3, and the installation and maintenance of the BDF target complex and experiment may become a limiting factor. The alternative of a second shaft is to be considered. 

While the ventilation in TT85-TDC85-TCC8 and ECN3 appears to be sufficient for the BDF, the section of TCC8 where the target station will be implemented should have a new dedicated ventilation system to confine the target area at an underpressure with respect to the rest of the facility. This ventilation partition should allow the maintenance of the target complex including the full operation of an overhead travelling crane. 

For any civil engineering considerations, the activation level of the tunnel concrete at the proposed BDF target complex location in TCC8 as well as in the downstream ECN3 cavern is expected to be low with mostly residual dose rates at the background level and a maximum dose rate at contact of 2\,$\mu$Sv/h at the former NA50 location.

The control cubicles of the ventilation of TCC8 and ECN3 are located in the room 1203 at the top of TCC8, just before the entrance to ECN3. With the BDF target complex located towards the end of TCC8, the compatibility between the electronics and the prompt radiation levels needs verification. Should it be necessary, the control cubicles may be moved to the surface or to the end of ECN3.

For the scheduling of activities in TCC8, the installation of the target complex may continue while the North Area is operated with beam, including the M2/EHN2 facility, currently hosting the COMPASS experiment.

\subsection{ECN3 - Experimental area}
\label{sec:ECN3_EA}

The ECN3 dimensions and infrastructure only require very limited modifications to host the muon shield and the SHiP detectors (see Fig.~\ref{fig:ECN3_views} in Appendix I). 

The current beam height is 2.8\,m above the ECN3 floor, leaving 6\,m of free space under the existing 45-tonne crane. The space is sufficient for the re-optimised experimental configuration. Below the beam height, the spectrometer section and a part of the decay volume are in conflict with the current floor level. To accommodate this, the cavern floor will be lowered by about 2.5\,m across the full width of the cavern and over a length of 25\,m, and a more shallow and narrow trench will be constructed under a part of the decay volume.  The floor construction and the ground conditions at this location are well known. The civil engineering works require the installation of micropiles and anchorages along the full-width excavation into the existing unconsolidated and consolidated moraine layer to underpin the walls of the tunnel and to maintain the stability of the structure. 

An initial study aiming at adapting the construction of the detectors to an underground location, accessed by a shaft, indicates that the crane capacity of 45\,tonnes is sufficient. Available information supports the possibility to add a second crane on the same rails, with lower capacity and a minimum distance between the cranes to be respected, in order to add more flexibility in the installation phase. 

The single shaft to TCC8 and ECN3 in building 911 has currently dimensions $4 \times 8$m$^2$, and is served by a 40-tonne crane. At the bottom, the access to ECN3 is via a 15\,m tunnel section with a width and free height of $4.8\times 6$\,m$^2$. The adaptation of this shaft and tunnel into ECN3 needs detailed study, taking into account the requirements from the target complex and the experiment.  No underground structure or infrastructure are present on the south-west side of the shaft/bld 911, making it possible to envisage an enlargement of the shaft and the possibility of a second shaft.

The ECN3 experimental area is served by a large service building (918) with two personnel shafts leading down to ECN3. The service ducts between ECN3 and 918 are sufficient. The building 918 includes $\sim$380\,m$^2$ of rooms for control, meetings, and offices, $\sim$110\,m$^2$ for electronics racks and online systems, $\sim$70\,m$^2$ for labs, workshops, storage, $\sim$16\,m$^2$ for equipment related to electricity, cooling, ventilation, and $\sim$47\,m$^2$ for gas. With the exception of a larger workshop, this space is equivalent to what was defined for the service building in the CDS design, and may require only regular renovation. 

In view of the 8\,m thick mound of soil currently present on top of TCC8 and ECN3, as well the 5\,m thick and 40-45\,m wide back-fill up to 100\,m downstream of ECN3, the prompt radiation levels in the proximity of the facility from neutrons, and to some extent also from muons, are expected to be significantly lower than at the TT90-TCC9-ECN4 site without further increasing the level of back-fill. With the horizontal muon deﬂection, the vast majority of high-energy muons are well contained under the current ground levels. Taking into account the location of the target with respect to the CERN boundaries and the local topology, it is expected that the optimisation dose goal of less than 10\,$\mu$Sv/y to members of the public~\cite{SafetyCodeF} can be reached.

The ECN3 experimental area has no assembly hall. The plan is to identify alternative existing assembly halls for the BDF. Several options are possible at the North Area, and the access road to ECN3 is adequate for the required transports. The extension of the EHN1 currently hosting the Neutrino Platform is suitable for the pre-assembly of all large detector components. As in all options, the plan will require further discussion on the availability of the existing halls.

\section{TNC option}
\label{TNC}

TNC is currently hosting HiRadMat as its exclusive user. The facility occupies about \SI{60}{m} of the first part of the hall up to the proton dump. A cool-down area for the experiments running in HiRadMat extends until the end of TNC. The facility will be operated at least throughout all of Run~3.  A study group is actively investigating interest in the HiRadMat facility after LS3 and possible upgrades, aiming at producing a proposal by end of 2022. The study considers compatibility with new users of the zone. 

\subsection{TNC - Beam delivery}

The slow extraction equipment in LSS6 has been largely decommissioned since the West Area stopped operation about 20 years ago in order to make way for a dedicated fast extraction system to serve the LHC. Although there are several machine protection concerns regarding the fast extraction of the bright, high intensity LHC beams in close proximity to the fragile wire arrays in the electrostatic septa, the re-implementation of a slow extraction system is feasible. A redesign of the extraction region would be necessary~\cite{Goddard:448332, Goddard:555902}, including the necessary beam instrumentation. In addition to the re-installation of the electrostatic septa, new low impedance, large aperture fast extraction magnets with a larger deflection angle will need designing and installing to safely extract the LHC beam past the electrostatic septa. A dedicated R\&D study would be needed to design and estimate the cost of the upgraded fast extraction magnets with today's stringent SPS impedance requirements in mind. The thin magnetic septa are still installed but their power supplies will need to pulse for much longer than is presently the case for LHC fast extractions. Electrostatic septa and their ancillary systems will need to be reinstalled. Although the recent R\&D effort has improved the ease of operability of the SPS slow extraction system, e.g. beam loss and induced radioactivity, machine reproducibility and spill quality, etc., it should be noted that the increased operational overhead of running two slow extraction systems needs to be a consideration, including creation of another high radiation area inside the SPS tunnel.

Beam transport from the SPS will follow the existing TT60 and TT66 beam lines, currently used for the transport of the HiRadMat beam. Recent studies of the local geometry shows that a total of around \SI{70}{m} of bending magnets will need installing to bring the beam trajectory in the horizontal plane and towards a direction which is compatible with the civil engineering of the target complex and the experimental area in the TNC. 
This concept does not modify the existing LHC extraction line but maintains the beam line at a level of \SI{410}{masl}, which requires excavating a flat trench starting at the entrance of the TNC tunnel. Using the compact dilution system and a 50\,m vacuum chamber, the target will be located $\sim$70\,m from the BA7 access tunnel (TJ7).

\subsection{TNC - Target complex}

The location at a depth of 35\,m makes the TNC particularly suitable from the point of view of radiological considerations. As TT90-TCC9-ECN4 and ECN3, the TNC is entirely located on the CERN domain. Constructing the target complex and the experimental area in the horizontal plane  
allows safe containment of the high energy muon flux over a distance of more than a kilometre downstream of the facility. The underground location also has a signiﬁcant impact on the amount of shielding required laterally around the target system. This is because TNC is located within the impermeable molasse without contact with shallow aquifers, and the impermeable molasse is unsuitable for drinking water exploitation. 

In TNC, the HiRadMat facility will need to be entirely dismantled. This includes the three irradiation stands, the beam dump and its associated shielding. The current HiRadMat dump shielding comes from the WANF T9 target shielding and is still very radioactive.

From similar past experience, the dismantling of the beam line and the associated services requires 4-6 months, following 6-8 months cool-down time. This includes activities throughout the whole TNC tunnel, including the storage of iron blocks and the highly radioactive copper collimator from the WANF dismantling, with a large gradient of dose rate. 

The shielding surrounding the BDF target will be reduced with respect to the TT90-TCC9-ECN4 configuration and will be embedded in a nitrogen confinement (see Section~\ref{sec:ECN3_TC} and Fig.~\ref{fig:Target_complex_TCC8_integration} in Appendix I). The size of the future cavern for the target complex is to some extent constrained on one side by the proximity of the TT61 tunnel. The tunnel leads to the West Area and is required to transport beam-line equipment to TCC6 and the nearby TI2 transfer line.

The maintenance of the target system will be performed with a modified configuration of the trolley concept developed in the CDS design. This consists in having all the active components of the target complex mounted on a trolley, designed to be extracted upstream along the beam line, where the equipment will be accessible with the help of a new under-hung crane. As in all options considered, a fully remote handling maintenance scheme with the support of telemanipulation and robotics is required.

There is no existing building near to BA7 capable of housing all the dedicated services for the BDF target complex, such as the new cooling and ventilation equipment. Hence, a dedicated auxiliary building will be required. The new building will also house the target system ancillaries, such as the buffer zone for the temporary storage of spent equipment.

The PA7 access shaft will be used as the principal channel for all the services required for the target station. The delivery of all the component for the installation, maintenance, and decommissioning of the target complex will also use the existing PA7 shaft. Due to the size, carrying capacity and accessibility of the goods lift (25-tonne carrying capacity with an internal cabin size of 6\,m deep, 2\,m wide, and 2.35\,m high), the foreseen size and weight of the equipment of the facility should be reduced compared to TT90-TCC9-ECN4 option.

The section of the new area housing the target station should have a dedicated ventilation system and be locally confined with underpressure with respect to the rest of the facility.
This ventilation partition should allow the maintenance of the target complex including the full operation of the under-hung crane.

As is currently the case for HiRadMat, access to the target complex and the beam line will be possible during SPS and LHC operation. The area is only inaccessible during the LHC injection phase. As a result, installation and commissioning may continue even after the restart of the injectors for Run 4.

\subsection{TNC - Experimental area}

With the limited dimensions of the TNC tunnel, the experimental area will need to be entirely excavated (Fig.~\ref{fig:TNC_views} in Appendix I) as a new volume, with dimensions that are similar to the dimensions of ECN3 with the limited modifications discussed in Section~\ref{sec:ECN3_EA} (Fig.~\ref{fig:ECN3_views} in Appendix I), together with additional space behind for the detector construction. As a result, a significant part of the existing TNC tunnel will need to be completely or partially demolished. With the optimised orientation of the whole underground complex in the horizontal plane, the experimental hall stretches past the end of the TNC, but with no interference with the WANF decay tunnel. Since there is limited space for a surface building immediately on top of the underground complex, additional space must be excavated to house infrastructure, detector services, and electronics.

The excavation is proposed to be carried out using the umbrella arch method by creating a temporary support system forming a structural umbrella prior the demolition of the existing tunnel to increase the excavation front sustainability and to minimise the risks associated with the deep excavation. Where the new underground facility only requires a partial demolition of the existing structure, the space between new tunnel and the parts of the existing tunnel left untouched will be filled with a suitable material.
The concrete of the whole TNC tunnel is still radioactive from the WANF operation with maximum dose rates at the location of the former WANF target station and the beginning of the decay tunnel of the order of 50\,$\mu$Sv/h. The dose rates in between the target station and the decay tunnel are mostly expected to be of the order of a few\,$\mu$Sv/h. The civil engineering works will therefore need to be planned and carried out accordingly. 

The significant volume of soil that will be excavated might have an impact on the existing infrastructures in the area. Therefore the works need to be carried out in a way that minimises the effect on the stability of the existing tunnels in the close vicinity of the construction site.

The current emergency escape shaft at the end of the TNC leading up to the surface hut 846 will be replaced by the construction of a 8-10\,m diameter standard shaft for the experiment equipment and personnel, together with a surface building for reception of equipment and intermediate storage. The shaft must also carry services between the surface and the detector. A short corresponding road and galleries will be constructed with direct connection to the SM18/BA7 area, where a service building is needed to house space for detector operation. A location for the building must be identified. The shaft and the road will be entirely on the CERN domain where there is currently no infrastructure.

As all the alternative locations, the TNC option relies on the use of existing surface halls for the detector pre-assembly. The road network at around BA7/SM18, and to Meyrin/Prevessin are suitable for this purpose.

\section{TCC4 option}
\label{TCC4}

TCC4 is currently hosting the AWAKE plasma acceleration R\&D project that is aiming to continue the development throughout Run~3 with the inclusion of an additional plasma cell. AWAKE has also proposed a cleanup of the entire TCC4 during LS3 in order to extend the setup to a scalable plasma acceleration facility, making use of the first $\sim$60\,m of TCC4 after LS3~\cite{AWAKE_CSreview}. The BDF implementation studies consider both a downstream location with a beam line bypassing the AWAKE facility, and an exclusive use of TCC4, depending on the decision on the future plans of AWAKE.

The situation of the TCC4, located at a depth of about 60\,m, means that the radiation protection aspects are similar to those of TNC.

\subsection{TCC4 - Beam delivery}

Slow extracted spills have never been performed through LSS4 and therefore no slow extraction equipment of any form is installed. A study would be needed, similar to the one carried out during the conception of the LHC extraction system in LSS6, to assess the feasibility and required hardware modifications to LSS4. A conventional slow extraction system is feasible but requires a complete redesign of LSS4, including the installation of electrostatic and thin magnetic septa, along with new large aperture fast extraction kickers for fast extractions to LHC. The phase advance of the presently installed extraction sextupoles will also need to be checked, and other new magnets and power converters may be required. On a longer timescale, a non-resonant extraction using thin bent crystals in LSS4 could be an interesting option to simplify and reduce the cost of providing slow extracted beams through LSS4. In this case the beam would be bumped slowly and directly into a crystal to push it into the extraction transfer line. Although a concerted effort is being made to develop highly efficient crystals for slow extraction, the timeline for the required technological development is far from certain.

The transport of the beam along the TT41 tunnel will use the existing beam line and all its elements, currently used for the transport of the proton beam to the AWAKE R\&D facility. Cohabitation with the AWAKE facility will need to be evaluated in detail. Branching off the existing line near the end of the TT41 tunnel, a new 50\,m bending section followed by a 50\,m compact dilution section will bring the beam in the horizontal plane but also result in a beam height approximately 1.85\,m above the existing TCC4 tunnel. Due to the strong downward slope (5.7\%) and limited dimensions of TCC4, the target complex and the experimental area will be excavated in a plane, which, for the larger part, is on the top of the existing tunnel. The TCC4 tunnel will have to be partially demolished and filled with concrete and earth to support the new structures, hence not taking advantage of the existing TCC4 tunnel.

Due to the extra complication of the demolition needed in TCC4, an alternative is to take the beam line off axis and excavate an entirely new volume underground for a short beam line section, the target complex and the experimental hall. 

\subsection{TCC4 - Target complex}

In TCC4, the remaining part of the CNGS target complex will need to be entirely dismantled. This includes the CNGS target, horn, and their associated shielding systems. From similar past experience, the dismantling of the beam line requires 12-20 months, if executed during Long Shutdown 3. This includes activities throughout the whole TCC4 tunnel with a large gradient of dose rate reaching up to 50\,mSv/h at contact for the target. 

The configuration, the implementation, and the maintenance scheme of the target complex will be similar to the TNC option. Target systems services such as control, cooling and ventilation will be located in a new cavern protected from the prompt radiation generated by the impact of the beam on the production target in order to avoid issues with radiation to electronics. The size of the future cavern for the target complex at this location is not limited by the close vicinity of other tunnels compared to the TNC option. A new overhead travelling crane will be required for the new cavern. Space underground will also be needed for the buffer zone for the temporary storage of used equipment.  

A major constraint with the current access to TCC4 is the slope in the access tunnel TAG41 ($\sim$5\%) and the transport limit of 7.5\,tonnes for standard transport and handling equipment. In addition, the length of this access tunnel of 800\,m and the cross section severely limit the size of components and the rate of use. The size and weight of the individual components of the facility would have to be significantly reduced compared to all the other options. Since the experiment requires a proper vertical material access shaft to the experimental hall, the alternative solution will be sharing the shaft for the installation and the maintenance of both the target complex and the experiment. This option removes the previous constraints, at the potential cost of having to increase further the size of the cavern to implement the transport path to the target complex from the common shaft. It also requires careful planning of the use of the common shaft during the installation phase.

As in all options, the section of the new tunnel where the target station will be implemented should have a new dedicated ventilation system to confine the target area with an underpressure with respect to the rest of the facility. The partitioning should allow the maintenance and the remote handing of the target complex with the help of an under-hung travelling crane.

\subsection{TCC4 - Experimental area}

As discussed above, the new underground areas will require demolishing a part of the tunnel crown and excavate a volume of soil on top of the TCC4, with dimensions that are similar to the dimensions of ECN3 with the limited modifications discussed in Section~\ref{sec:ECN3_EA}. Furthermore, as for TNC, additional space is needed to house underground the infrastructure and detector services, and electronics. Taking into consideration the previous use of TCC4, the existing concrete and the surrounding soil is highly activated with dose rates in the former CNGS target/horn and reflector regions expected to still reach up to 300\,$\mu$Sv/h and 60\,$\mu$Sv/h in 2026, respectively. The civil engineering works will therefore need to be planned and carried out accordingly. To evacuate the excavated material, a civil engineering shaft will be required on top of the underground facility during the execution of works. Once the works are completed the shaft will be turned into an access shaft for the experiment.

The installation, maintenance and decommissioning of the experiment will make use of the civil engineering shaft refurbished into a standard shaft for personnel and material. The shaft requires an associated surface building for reception of equipment and intermediate storage, and an adequate road connection for transport of heavy components.  As in all options, use of TCC4 relies on using existing surface buildings on the Meyrin and Prevessin sites for detector pre-assembly. The location has currently no CERN or public surface infrastructure.  As it is outside of the CERN domain and just at the border of the natural reserve Mategnin Les Cr\^ets, the TCC4 option requires very careful evaluation. If the current level of general services (power, water, network etc) are not sufficient, they will need to be routed from the SPS BA4.

\section{Comparative summary and outlook}

The assessments of the ECN3, TNC, TCC4 options, as well as the revision of the TT90-TCC9-ECN4 option, rest on the requirement that the physics scope and the physics reach of the BDF/SHiP proposal are preserved. The motivation behind the investment in a new or upgraded facility stem from the considerable physics potential that may be unlocked by exploiting the substantial yield of protons that is currently unused at the SPS. A comparative summary of the main aspects of the BDF options follows (for details, see the specific sections).
The comparison elucidates the considerable cost reductions possible in the alternative locations with respect to the CDS version of the TT90-TCC9-ECN4 option (Table~\ref{tab:cost-summary-material-overview}). \\

The TT90-TCC9-ECN4 remains a valid option. The recent progress on optimising the design (Section~\ref{sec:TT90optim}) has an important impact on its cost, and reduces some risks and uncertainties identified during the CDS study. No effort has been made yet to re-estimate the cost of the revised design.\\

All the alternative locations can be made compatible with the BDF/SHiP requirements and constraints, although at very different levels of civil engineering and technical
modifications, and hence cost. The integration and the installation scenarios in a fully underground location are more complex, but no showstopper has been identified. All the alternative options, as well as the re-optimised TT90-TCC9-ECN4 option, now rely on the use of existing assembly halls on the CERN sites instead of a dedicated on-site building.\\

\noindent
{\bf \underline{Civil engineering and radiation protection}}

The {\bf ECN3} option requires very limited modifications of the existing ECN3/TCC8 underground complex and surface site. The ECN3 floor needs lowering by about 2.5\,m across the full width of the hall over a length of about 25\,m, together with a trench under a part of the detector, and the access shaft common to TCC8 and ECN3, with its associated access building, is likely to require modification. No existing infrastructure interferes with these works. At the foreseen location of the target system, the cavern size is sufficient but the floor needs modification to incorporate shielding and route services. The activation level of the tunnel concrete at this location is expected to be low with mostly residual dose rates at the background level. Consequently, this option comes with significantly lower risks and uncertainties compared to the TT90-TCC9-ECN4 option, and the other alternative options, making estimation of the implementation cost straightforward.  

With the specificities of the BDF implementation and operation, radiation protection constraints associated with the beam line, the target complex and the experimental area in the ECN3 option appear solvable with acceptable measures. The existing back-filling on top and behind the ECN3 hall is expected to improve on the stray radiation even compared to the TT90-TCC9-ECN4 option. The flux of muons emanating from the muon shield appears sufficiently well contained within the ground over the distance required to range out the muon energies, and to respect the dose constraints for members of the public.\\

The depth and location of the {\bf TNC} option on the CERN domain make it particularly suitable from a radiation protection point of view. The proximity to the CERN infrastructure, general services, and the underground access at the BA7/SM18 area are relatively well adapted to the needs of the BDF. However, TNC involves considerable underground civil engineering works consisting of the destruction of most of the TNC tunnel and excavation of a hall similar in size to ECN3 with a shaft for the experiment located at the end of the hall. The demolition of the TNC tunnel leads to large volumes of activated concrete. As opposed to TT90-TCC9-ECN4, where this material can be used as backfill at strategic locations, the construction at the TNC does not include similar needs for backfilling. Additional surface buildings are needed on the BA7/SM18 sites to house services and operational space for the target complex and the experiment.\\

The {\bf TCC4} option is similar to TNC but with additional challenges not present at the other locations. TCC4 has no existing direct material shaft, neither surface buildings. It only has a very limited material and personnel access from BA4. The location outside of the CERN domain, and the proximity to the natural reserve Mategnin Les Cr\^ets, put severe restrictions on the construction of even a minimal surface site and a road network to transport components.  In addition, the strong downward slope in TCC4 leads to very limited re-use of the existing underground complex. Practically, the entire volume required for the target complex and the experimental hall, as well as underground space for the target complex and detector services, must be excavated as a new volume, with the additional constraint of managing the activated concrete from the partial demolition of the TCC4 tunnel. \\

\noindent
{\bf \underline{Beam delivery}}

Significant progress has been made in recent years to implement new beam loss reduction techniques for the SPS slow extraction by improving the efficiency of the existing electrostatic septum system in LSS2, which serves the North Area and the {\bf TT90-TCC9-ECN4} and the {\bf ECN3} options. When considering locating the BDF at another extraction point in the SPS, as for the {\bf TNC} and {\bf TCC4} options, one must include the cost of implementing a second system in the SPS, alongside a significant cost of maintenance and the increased operational overhead of running two slow extraction systems. \\

The delivery of dedicated BDF spills to {\bf ECN3} requires relatively minor BDF-specific modifications of the beam lines to guarantee the loss-free transport of the BDF spills. A dedicated BDF optics will need implementing in TT20 to avoid beam splitting and hardware modifications will be required to facilitate the vertical bypass of the T4 target. The function of the T4 target on the long fixed-target spill cycles will be preserved. The result is that the implementation of the BDF in ECN3 involves no interference with the function of the other North Area facilities. General modifications to consolidate the beam lines to the North Area such as renewed instrumentation and beam intercepting devices are already considered in the context of the North Area Consolidation programme. For that reason, a timely evaluation of the future beam requirements, including accommodating higher intensities, for the North Area Experiments is mandatory. \\

The beam line to the {\bf TNC} had a slow extraction system at the time of the WANF but it was decommissioned to ease the implementation of the fast extraction into TT60-TI2 for the LHC. The re-implementation of a slow extraction system is feasible but requires the redesign of the extraction region and R\&D to develop large-aperture and low-impedance fast extraction magnets to implement both the slow and fast extraction functions. From the TT60, the beam transport will be similar to the current beam line for HiRadMat with the additional constraint that the beam should be bent towards the northern side of the TNC such that the civil engineering of the facility is not interfering with the TT61 tunnel. The beam will also be brought into the horizontal plane before the envisaged location of the BDF target complex. No other changes are expected to implement the beam line.\\

For {\bf TCC4}, the existing beam line may be used as it is presently installed. However, it has never had a slow extraction system. Consequently, it requires a complete redesign of the SPS LSS4 with the installation of new slow extraction components. In this option, it is considered more attractive to pursue the possibility of non-resonant extraction with the help of thin bent crystals, but the success of this technique depends on future R\&D.\\

\noindent
{\bf \underline{Target complex and experimental area}}

The deeper underground location of all the alternative options has brought forward an updated design of the BDF target complex. With the additional natural shielding of the fully underground locations and the existing tunnels, and the longer time for short-lived isotopes to decay before the air is released into the environment, the shielding around the target can be reduced. While it is already the case for ECN3 compared to TT90-TCC9-ECN4, it is particularly true for the TNC and the TCC4 options as these are in the molasse. The inert gas in the vessel embedding the entire target shielding may also be changed from helium to nitrogen, simplifying the design of the vessel and of the respective purification system. The reduced shielding and the limited access from the top has further driven the design of the complex to implement access to the target system from the front and the side. As a result, the target complex may be incorporated into the existing TCC8 in the {\bf ECN3} option. It also requires less civil engineering in the {\bf TNC} and the {\bf TCC4} options than in the TT90-TCC9-ECN4 option. \\

In all the alternative options, additional new space is needed to house the target services, i.e. the dedicated ventilation, cooling and purification circuits for the target system and the inert gas. Space is also needed for handling tools, and spent components that are activated. At {\bf ECN3} and {\bf TNC}, these peripheral requirements will be integrated into a dedicated surface building, while additional space is needed underground in the {\bf TCC4} option. \\

For the {\bf ECN3} option, the dimensions of the {\bf ECN3} hall and the existing infrastructure (both above and underground) are sufficient to host the muon shield and the SHiP experiment with minimal civil engineering modifications. The space in the existing surface buildings is also sufficient to house the detector services and detector operation. In the {\bf TNC} option, a new surface building is required near to BA7/SM18. In {\bf TCC4}, the equivalent space must be split between space underground, and potentially a new building on the SPS BA4 site. \\


\noindent
{\bf \underline{Road map}}

The implementation of the BDF at TT90-TCC9-ECN4 and, in particular, at ECN3 is synergetic with the North Area Consolidation programme~\cite{NACONS, NACONSadd}, both in phase 1 and 2, for the primary beam lines (magnets, supports and alignment, power supplies, cabling, beam instrumentation), the TT20 splitter magnets and primary target areas, but also the renovation of civil structures. A timely decision on the future programmes allows scoping and optimising the consolidation programme for post-LS3 projects. \\

{\bf The implementation of the BDF at the alternative locations is primarily a re-optimisation of the facility and the detector. It builds on the concepts developed in the extensive joint studies performed during the six years of the TP and CDS phases, which concentrated a large part of the effort on tuning the design of the components to maximise the physics background suppression and the signal acceptance. Assuming that the hidden sector exploration is taken as a priority in the diversity physics programme of CERN, the Technical Design Report for the retained location could be delivered within two years of the project's approval. This is compatible with commencing the implementation of the BDF in any of the options during Long Shutdown 3. \\}

Below is a summary of the highest-priority studies necessary to address conceptually the principal technical feasibility issues identified in this assessment, the implementation in terms of integration and civil engineering, and to produce accurate estimates of the material cost and CERN personnel needs, together with a detailed road map. For the retained option, these location specific studies can be completed by the end of 2022 with the current resources in the BDF team.\\

\noindent
{\bf \underline{Next steps in the ECN3 option}}

\begin{itemize}
\setlength\itemsep{-0.2em}
\item Beam line studies focusing on the optics and instrumentation for loss-free beam transport to the BDF target without splitting and bypassing the T4 target, including associated radiation protection studies of accidental loss scenarios 
\item Studies and optimisation of the target complex system in TCC8, integration, access and maintenance strategy, including handling and storage of radioactive components
\item Studies of the residual radiation, air/He/N$_2$ activation and soil activation  
\item Studies of the radiation levels above the ground within the fenced and non-fenced CERN domain
\item Consolidate the re-optimisation of the muon shield, detector layout, and the background evaluation
\item Integration of the experiment, including the installation scheme
\item Investigate the capacity of available power converters and cooling water for the beam line, target complex and experimental area
\item Civil engineering studies for target complex and experiment, focusing on the modifications and the material access
\end{itemize}

\noindent
{\bf \underline{Next steps in the TNC option}}

\begin{itemize}
\setlength\itemsep{-0.2em}
\item Studies for the re-implementation of slow extraction through LSS6, including beam instrumentation
\item Studies and optimisation of the target complex system, integration, access and maintenance strategy, including handling and storage of radioactive components
\item Studies of the residual radiation and air/He/N$_2$ activation 
\item Consolidate the re-optimisation of the muon shield, detector layout, and the background evaluation
\item Integration of the experiment, including the installation scheme
\item Investigate the capacity of available power converters and cooling water for the beam line, target complex and experimental area
\item Civil engineering studies for target complex and the experiment
\item Special measures to continue installation/commissioning after restart of the LHC (TI2) for Run 4 
\end{itemize}

\noindent
{\bf \underline{Next steps in the TT90-TCC9-ECN4 option}}

\begin{itemize}
\setlength\itemsep{-0.2em}
\item Target complex optimisation using the lessons learnt from studying the fully underground implementation
\item Studies of the residual radiation, air/He/N$_2$ activation and soil activation  
\item Studies of the radiation levels above the ground within the fenced and non-fenced CERN domain
\item Consolidate the re-optimisation of the muon shield, detector layout, and the background evaluation
\item Integration of the experiment, including the installation scheme, focusing on the reduced size of the underground cavern, single access point, and use of existing detector assembly buildings
\item Update of the civil engineering with the revision of the target complex and the experimental area, focusing on an optimisation of the engineering techniques
\end{itemize}

\noindent
{\bf \underline{TCC4}} requires studies that are similar to the TNC option. However, with the specific drawbacks of this option compared to the others, it is suggested not to pursue it.


\section*{Appendix I - Conceptual layouts}
\addcontentsline{toc}{section}{Appendix I - Conceptual layouts}

Preliminary conceptual layouts at the different locations.\\

\noindent
{\bf \underline{New extraction tunnel in TT90-TCC9-ECN4 option}}

\begin{figure}[!htpb]
\centering
\includegraphics[width=0.99\linewidth]{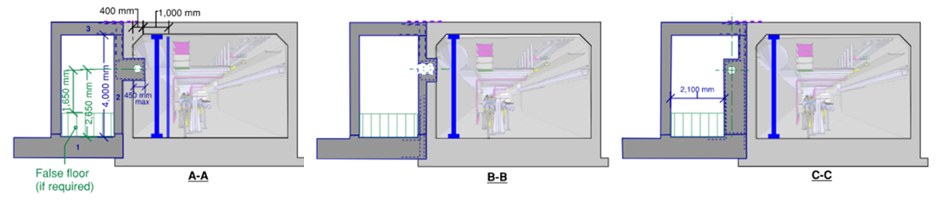}
\caption{TT90 wall penetration option details}
\label{fig:TT90_option_Details}
\end{figure}

\begin{figure}[!htpb]
\centering
\includegraphics[width=0.75\linewidth]{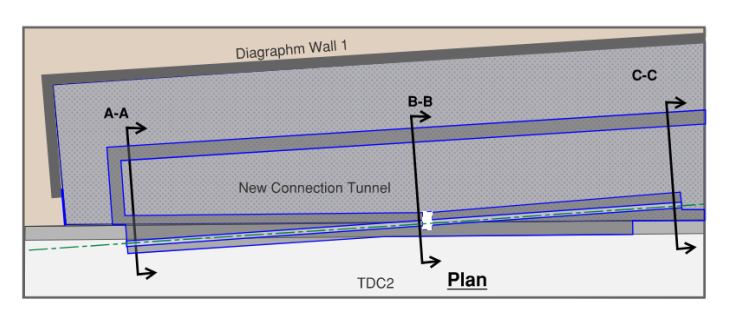}
\caption{TT90 wall penetration option plan view.}
\label{fig:TT90_Wall_penetration_option_Plan_view}
\end{figure}

\newpage

\noindent
{\bf \underline{ECN3 option}}

 \mbox{} \begin{center}
    \begin{sideways}
         \begin{minipage}{0.85\textheight}
                    \includegraphics[width=0.85\textheight,keepaspectratio]{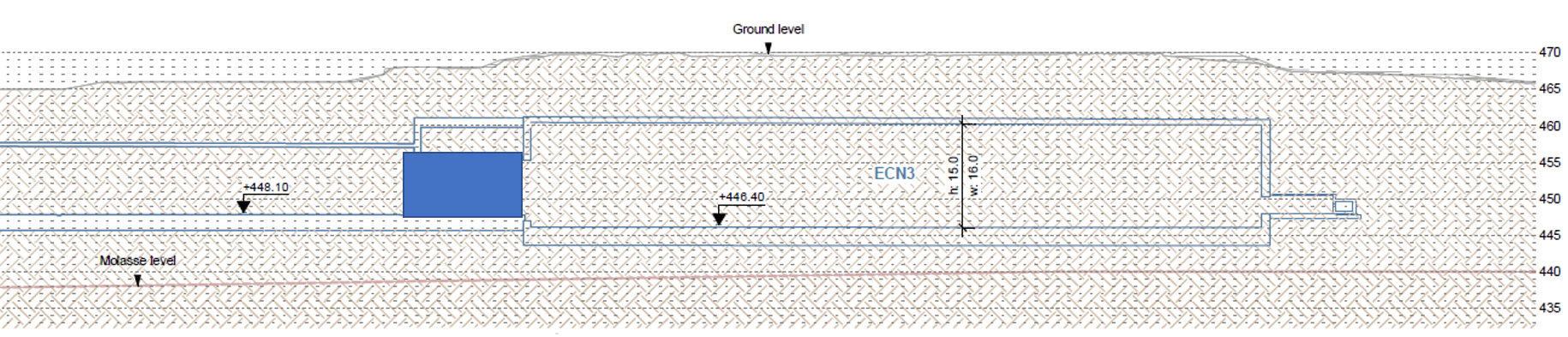}
                    \includegraphics[width=0.85\textheight,keepaspectratio]{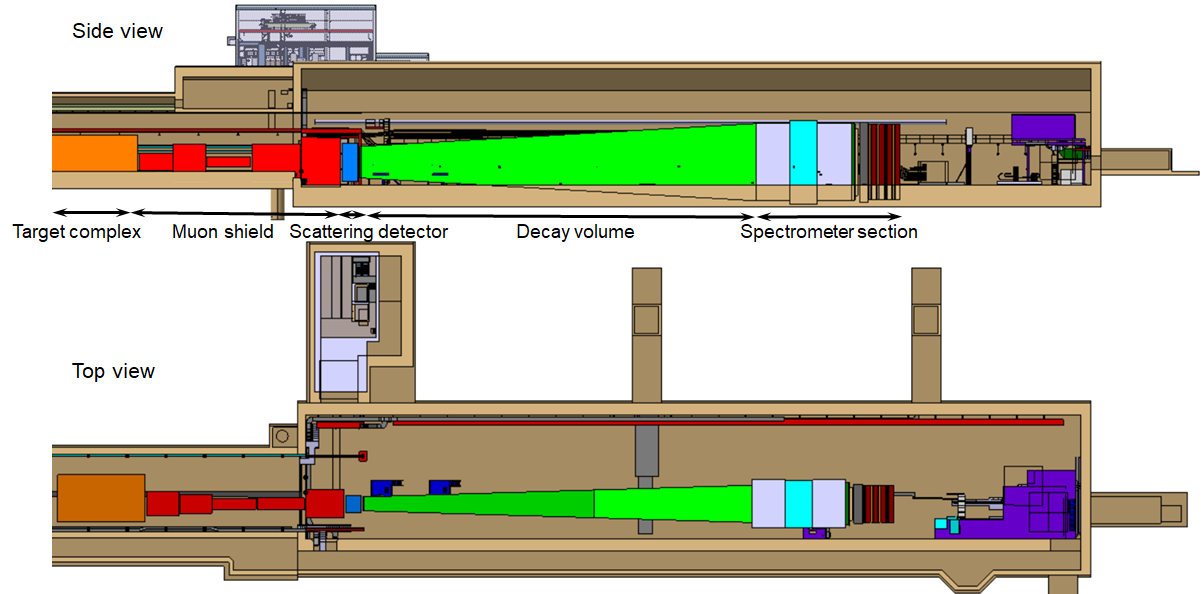}
                    \captionof{figure}{Views of the ECN3 option. Top view shows the ground profile along the relevant section of TCC8 and ECN3. The two bottom views shows a preliminary integration of the SHiP detector.}
         \vspace{0.2cm}
         \label{fig:ECN3_views}
         \end{minipage}
    \end{sideways}
    \end{center}
    
\newpage

\noindent
{\bf \underline{Updated target complex design as envisaged for the ECN3 option.}}\\

\begin{figure}[!htbp]

\centering
\includegraphics[width=0.75\linewidth]{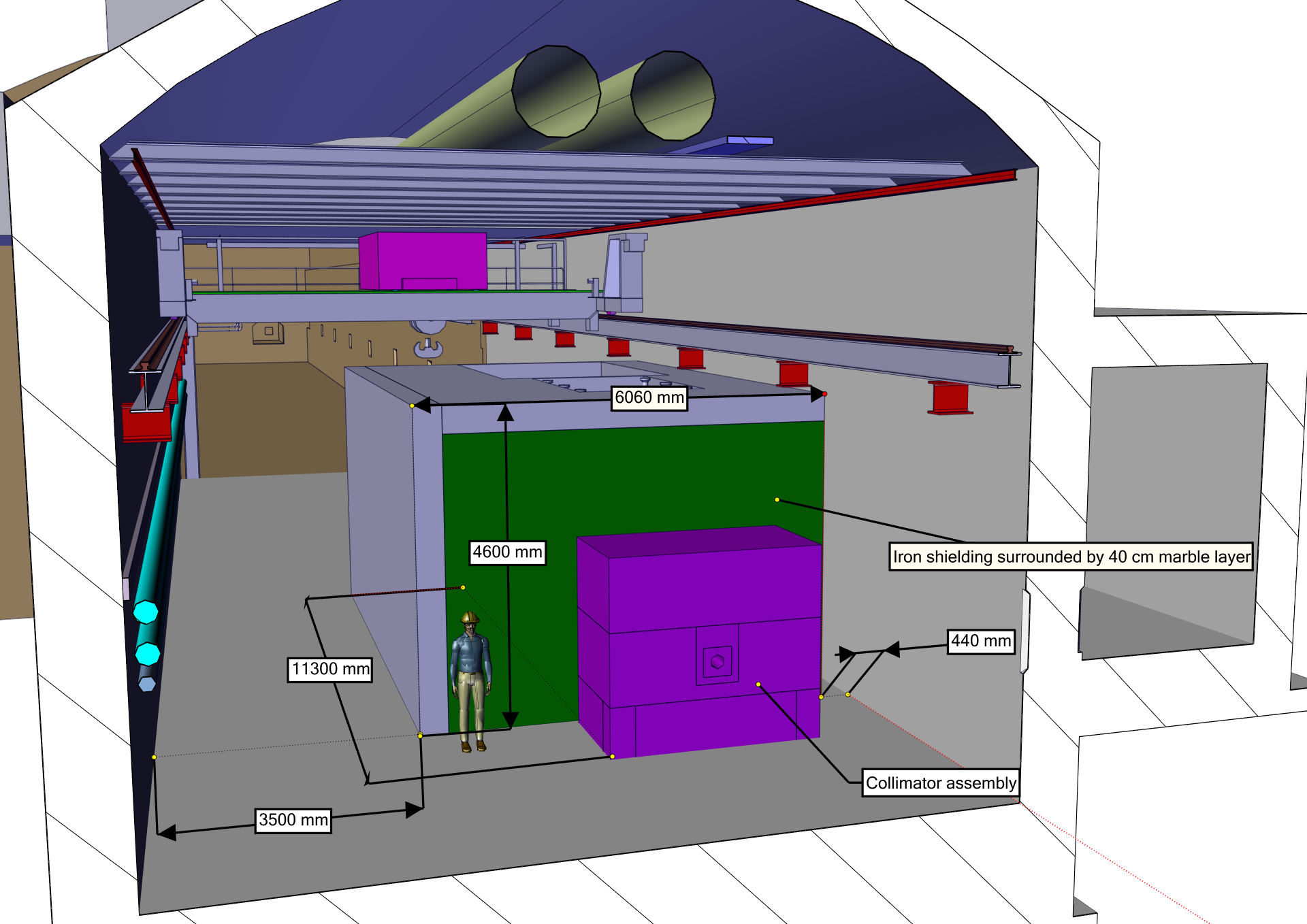}\\
\vspace{0.5cm}
\includegraphics[width=0.75\linewidth]{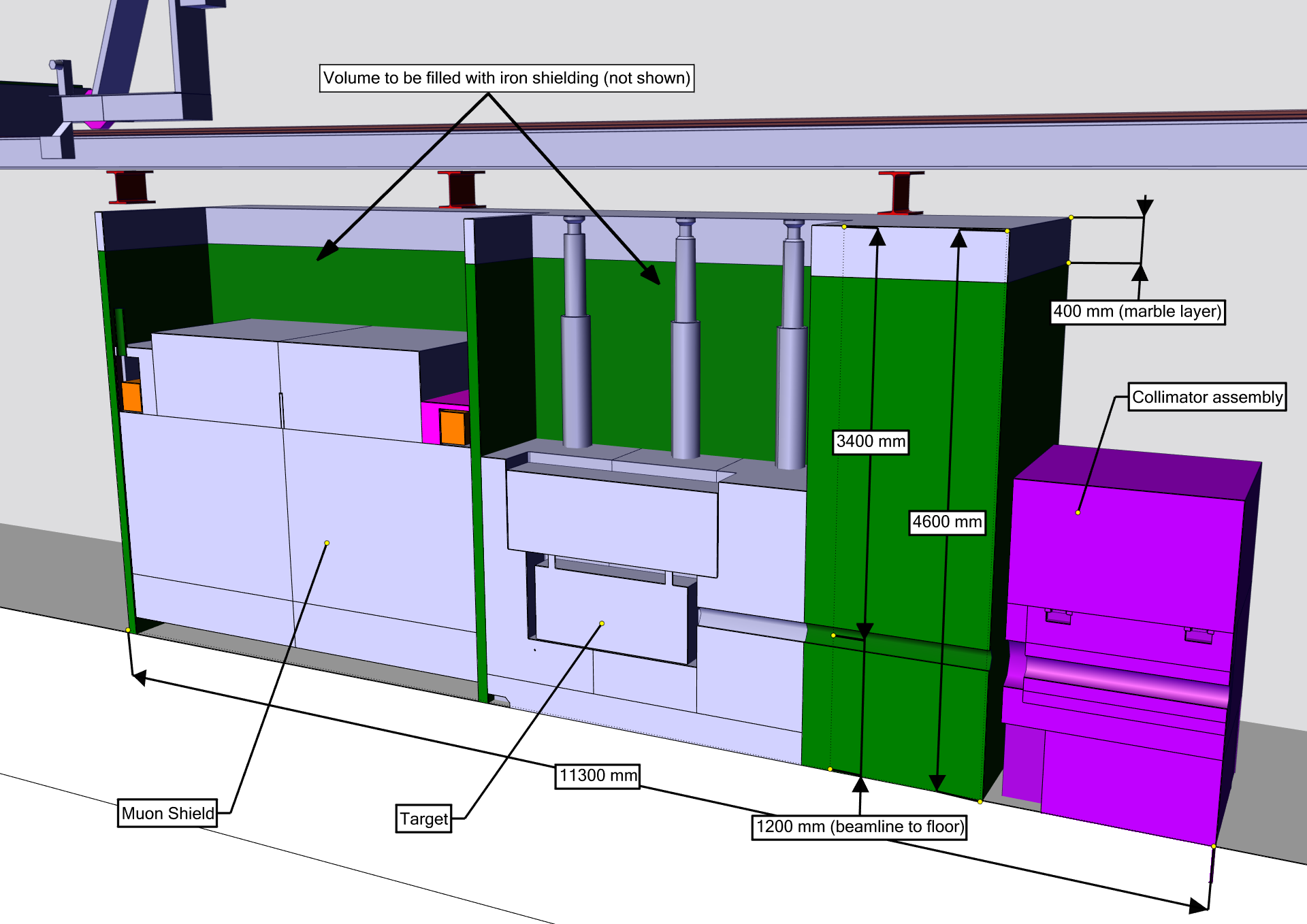}
\caption{The figures shows the conceptual design of the BDF target complex implementation in TCC8. The target is located within a water-cooled proximity shielding, in turn surrounded by cast iron and an external layer of marble. The target complex includes the magnetised iron shielding downstream of the production target. Similar designs may be envisaged in the TNC and the TCC4 options.}
\label{fig:Target_complex_TCC8_integration}
\end{figure}

\newpage
\noindent
{\bf \underline{TNC option}}
 \mbox{} \begin{center}
    \begin{sideways}
         \begin{minipage}{0.95\textheight}
                    \includegraphics[width=0.95\textheight,keepaspectratio]{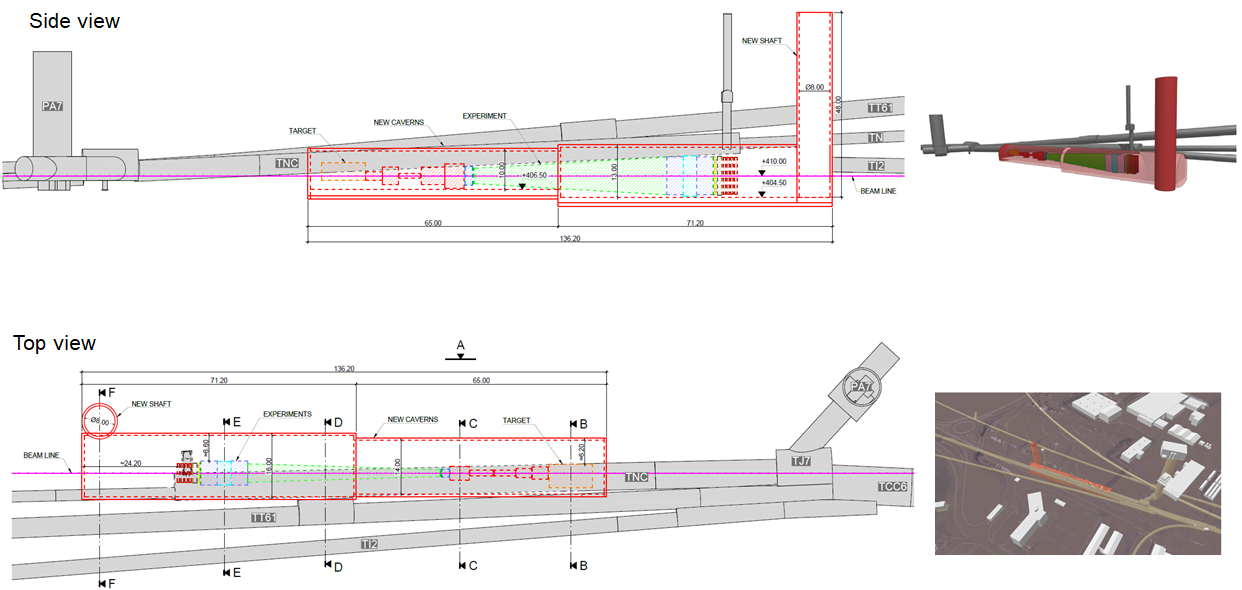}
                    \\captionof{figure}{Views of the TNC option. Note that the top and bottom figures are views along opposite directions.}
         \vspace{0.2cm}
         \label{fig:TNC_views}
         \end{minipage}
    \end{sideways}
    \end{center}


%
    
\newpage

\section*{Appendix II - Glossary of acronyms used in this document}
\addcontentsline{toc}{section}{Appendix II - Glossary of acronyms used in this document}

Map for further details: https://gis.cern.ch/gisportal/ (login required)
\begin{itemize}
\setlength\itemsep{-0.2em}
\item BA (1-7): Access buildings to the SPS tunnel
\item PA(1-7)  Access pits to the SPS tunnel
\item LSS(1-6): Long Straight Sections of the SPS accelerator
\item TT20: Tunnel for the transfer line between SPS and the North Area
\item TT41: Tunnel for the transfer line between SPS and TCC4
\item TT61: tunnel for the former beam line between the SPS junction cavern at BA6 and the West Area experimental hall
\item TT83: Tunnel shared by beam lines to ECN3 and EHN2
\item TT85-TDC85: Tunnels for the proton beam line to ECN3
\item TI2: Tunnel for injection line between SPS and LHC
\item P42 : Name of the proton beam line to ECN3
\item K12: Name of the kaon beam line for ECN3
\item M2: Name of the muon beam line for EHN2
\item P6: Former connection between the M2 and P42 beam lines
\item TDC2: Tunnel with the splitters of the beam lines for the North Area
\item TCC2: Underground hall with targets for the North Area
\item TCC4: Underground hall for the former CNGS target and magnetic horns, currently hosting the AWAKE project
\item TCC8: Underground target hall for the ECN3 experimental area
\item TNC: Underground hall for the former West Area Neutrino Facility target and magnetic horns, currently hosting HiRadMat facility
\item T4: Primary production target for the H6 and H8 beam lines
\item T10:  Primary production target for ECN3
\item TAX: Dump/absorbers after primary production targets
\item EHN1: Principal experimental hall (surface) on the North Area for the H2,H4,H6,H8 areas
\item EHN2: Experimental hall (surface) currently hosting the COMPASS experiment
\item ECN3: Experimental hall (underground) currently hosting the NA62 experiment
\item TT90-TCC9-ECN4: transfer line, target complex, and experimental hall for the proposed CDS design of the BDF
\item CDS: Comprehensive Design Study of BDF/SHiP
\end{itemize}

\addcontentsline{toc}{section}{References}
\bibliographystyle{JHEPx} %
\bibliography{references}%


\end{document}